\documentclass[twocolumn,amsmath,amssymb,superscriptaddress,aps,pra,a4paper,floatfix
]{revtex4-1}

\usepackage{graphicx}% Include figure files
\usepackage{dcolumn}% Align table columns on decimal point
\usepackage{bm}% bold math
\usepackage{gensymb}
\usepackage{xr}
\makeatletter
\newcommand*{\addFileDependency}[1]{% argument=file name and extension
  \typeout{(#1)}
  \@addtofilelist{#1}
  \IfFileExists{#1}{}{\typeout{No file #1.}}
}
\makeatother

\newcommand*{\myexternaldocument}[1]{%
    \externaldocument{#1}%
    \addFileDependency{#1.tex}%
    \addFileDependency{#1.aux}%
}
\myexternaldocument{supp}
\usepackage{mathtools, braket}
\usepackage{xcolor}
\usepackage[ruled,vlined]{algorithm2e}
\usepackage[noend]{algpseudocode}

\SetCommentSty{mycommfont}
\begin{document}

\title{Single-photon-level sub-Doppler pump-probe spectroscopy of rubidium}% Force line breaks with \\
\author{Paul %``Nighthawk" 
Burdekin}
\affiliation{Centre for Cold Matter, Blackett Laboratory, Imperial College London, Prince Consort Road, SW7 2AZ London, United Kingdom}
\author{Samuele Grandi}
\affiliation{ICFO -- Institut de Ciencies Fotoniques, The Barcelona Institute of Technology, Mediterranean Technology Park, 08860 Castelldefels (Barcelona), Spain}
\author{Rielly Newbold}
\affiliation{Centre for Cold Matter, Blackett Laboratory, Imperial College London, Prince Consort Road, SW7 2AZ London, United Kingdom}
\author{Rowan A. Hoggarth}
\affiliation{Centre for Cold Matter, Blackett Laboratory, Imperial College London, Prince Consort Road, SW7 2AZ London, United Kingdom}
\author{Kyle D. Major}
\affiliation{Centre for Cold Matter, Blackett Laboratory, Imperial College London, Prince Consort Road, SW7 2AZ London, United Kingdom}
\author{Alex S. Clark}
\affiliation{Centre for Cold Matter, Blackett Laboratory, Imperial College London, Prince Consort Road, SW7 2AZ London, United Kingdom}

\date{\today}% It is always \today, today,

\begin{abstract}
We propose and demonstrate pump-probe spectroscopy of rubidium absorption which reveals the sub-Doppler hyperfine structure of the $^{5}$S$_{1/2} \leftrightarrow$ $^{5}$P$_{3/2}$ (D2) transitions. The counter propagating pump and probe lasers are independently tunable in frequency, with the probe operating at the single-photon-level. The two-dimensional spectrum measured as the laser frequencies are scanned shows fluorescence, Doppler-broadened absorption dips and sub-Doppler features. The detuning between the pump and probe lasers allows compensation of the Doppler shift for all atomic velocities in the room temperature vapor, meaning we observe sub-Doppler features for all atoms in the beam. We detail a theoretical model of the system which incorporates fluorescence, saturation effects and optical pumping and compare this with the measured spectrum, finding a mean absolute percentage error of 4.17\%. In the future this technique could assist in frequency stabilization of lasers, and the single-photon-level probe could be replaced by a single photon source. 
%and observe spectral features not before seen. 
\end{abstract}

\maketitle

\section{\label{sec:level1}Introduction}
Vapours of alkali atoms have historically been an attractive system for studying light-matter interactions with a host of  applications including laser locking \cite{Pearman2002}, compact magnetometry \cite{Schwindt2004} and accelerometers \cite{Kasevich1991}. In some of the more recent work these atoms are being used for the storage of quantum states in the form of quantum memories \cite{Heshami2016}.
Atomic alkali vapour are promising candidates for these applications due to their large light-matter coupling \cite{Siddons2008a}, long-lived ground state coherence, and high room-temperature vapour pressures which result in large optical depths \cite{Hammerer2010}. There are also a broad range of quantum memory protocols designed and demonstrated for such systems \cite{Hsiao2018, Saglamyurek2019, Guo2019, Finkelstein2018, Hosseini2011}.  Many examples of light pulses containing multiple photons have been successfully stored in atomic ensembles at a variety of temperatures, along with the generation of entanglement \cite{Ding2015, VanLeent2019} and telecommunication wavelength conversion \cite{Radnaev2010}. However there have been relatively few demonstrations of single photon states interacting with atomic ensembles \cite{Namazi2017, Saglamyurek2019, Guo2019}.
Showing the interaction of single photons with the hyperfine levels of an atomic alkali vapour is typically performed in a magneto-optical trap (MOT), in order to remove thermally-induced broadening.  While techniques exist to resolve the sub-Doppler features of room temperature ensembles, traditionally these techniques use relatively high probe laser intensities \cite{Himsworth2010, Singh2010, Mohapatra2007}, compared with single photon emitters.

Here we demonstrate a modified sub-Doppler spectroscopy technique on a warm vapour of rubidium with a bright pump laser and a separate single-photon-level probe, which are counter-propating and independently tunable in frequency. The probe reveals the hyperfine spectrum of $^{87}$Rb and $^{85}$Rb. The different pump and probe frequencies not only cancel Doppler-shifts for atoms with zero longitudinal velocity, as is the case with standard saturated absorption spectroscopy, but also for all other velocities, which we find by plotting a two-dimensional spectrum. This technique can be useful for analysing the properties of atomic gases, laser locking, and can readily extended to replacing the single-photon level probe with a tunable, solid-state single photon source. 

\section{Experiment}

\subsection{Setup}
\begin{figure}
    \includegraphics[width=\linewidth]{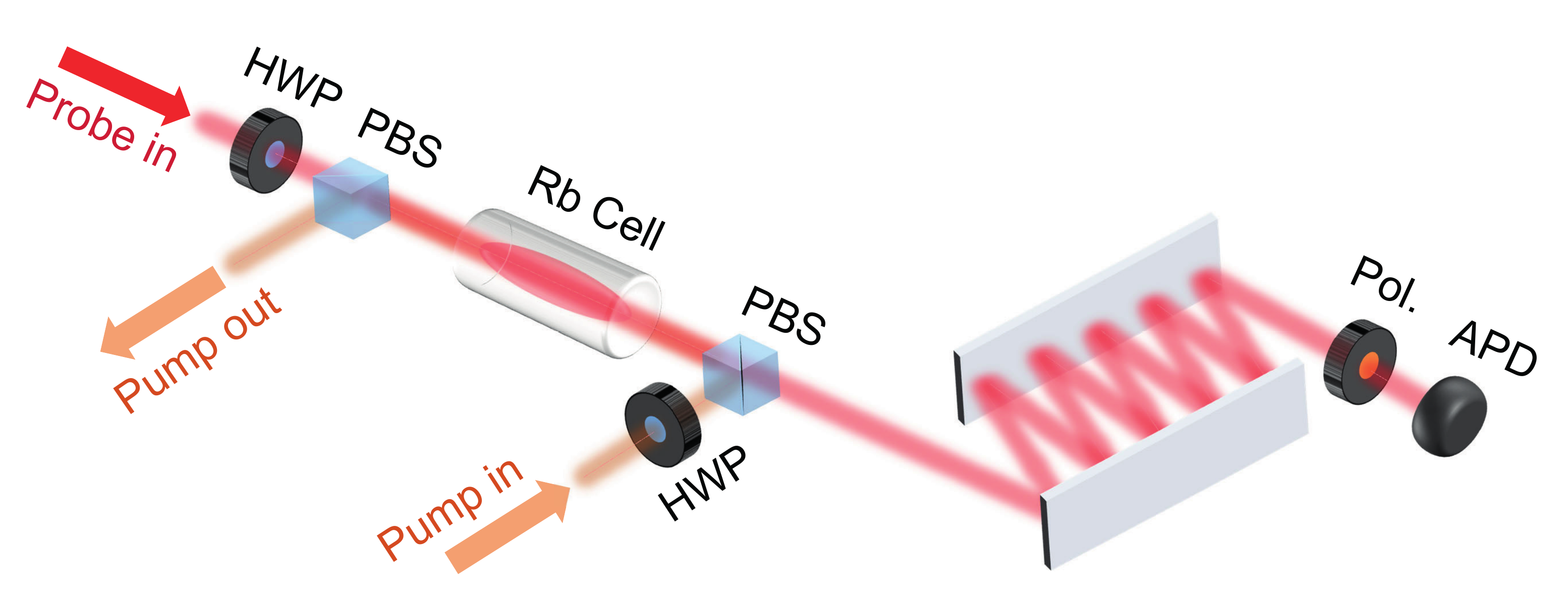}
    \caption{Experimental setup used to collect data shown in Fig.(\ref{data}).  The polarizations of the probe and pump beams are each rotated with a half wave plate (HWP) such that they are transmitted and reflected, respectively, by each polarizing beam splitter (PBS). After passing through the cell, the probe beam is reflected between two mirrors to increase the distance travelled to the avalanche photodiode (APD).  This reduces the acceptance solid angle of any fluorescence generated by the pump beam, which when combined with a polarization filter (Pol), aligned parallel with the polarization of the probe beam, greatly decreases the detected fluorescence.}
    \label{setup}
\end{figure}

A schematic of the experimental setup is shown in Fig.\,\ref{setup}.  An attenuated Ti:Sapphire laser (MSquared SolsTiS) was the single-photon-level probe, with the pump light being supplied by another Ti:Sapphire laser (Coherent MBR-110).  The polarization of the probe beam was set by a half-wave plate (HWP) to pass through the left-most polarizing beam-splitter (PBS), such that it passed through the vapour cell, interacted with the atoms, and passed through the right-most PBS. The polarization of the pump beam was set using a half-wave plate such that it was reflected by the right-most polarizing beam splitter (PBS) before passing through the cell in the opposite direction and orthogonal polarization to the probe beam, before reflection from the left-most PBS. The output probe then propagated for approximately two meters by bouncing back and forth between two mirrors to reject fluorescence emitted by atoms in the cell. It then passed through a final polariser, which was used to reject reflected pump light, before being collected in a multimode fiber and detected on a silicon avalanche photo-diode (APD: Excelitas, SPCM-AQR-14-FC). A temperature controller (Thorlabs TC200) was used to regulate the temperature of the 7.5\,cm long Rb vapour cell to $40\degree$C, which contained the natural fractional abundances of $^{85}$Rb ($0.7217$) and $^{87}$Rb ($0.2783$)\cite{Steck2003, Steck2013}.
 
The frequencies of the pump and probe lasers were independently tuned while the absorption of the probe was monitored on the APD.  Both lasers were held at a constant power before entering the vapour cell, using analogue PID controllers (SIM960) acting on acousto-optic modulators, with the probe locked to $2.5(1)\times10^6$\,photons/s arriving at the cell and the pump set to $791(2)\,\mu$W before the cell. No protection from stray magnetic fields was employed.

\subsection{Model}

%Intensity measured at output, components
%Work out cross section as function of pump
%Fluorescence term

To model the spectroscopy setup, we solved the Einstein rate equations for all energy levels in Rb, and for the two isotopes present in the cell. This simple model is suitable because we consider low laser powers, such that nonlinear optical effects may be discounted, and long time-scales, which cause coherent effects to be averaged out \cite{Himsworth2010}. Furthermore, spontaneous decay to `dark' ground states prevents any noticeable coherent effects \cite{Choi2015}.

The intensity measured at the detector is
\begin{equation}
    I(\omega, \omega_{p}) = I_\text{probe}(\omega, \omega_{p}) + I_\text{fluo}(\omega_{p}) + I_\text{BG} \,
    \label{eq:I}
\end{equation}
where $\omega$ is the frequency of the probe beam and $\omega_{p}$ is the frequency of the pump beam. The probe laser with input intensity $I_0$ experiences absorption in the cell according to the Beer-Lambert law, so the intensity at the detector is $I_\text{probe}(\omega, \omega_{P}) = \eta_\text{probe} e^{-N_{v}\sigma(\omega, \omega_{p}) I_{0} l}$ where $\eta_\text{probe}$ is the total efficiency for collection and detection of the probe beam,  $N_{v}$ is the number density of Rb, $\sigma(\omega, \omega_{p}$) is the absorption cross-section, and $l$ is the length of the cell. The intensity contribution from fluorescence caused by the pump beam is $I_\text{fluo}(\omega_{p})$, while $I_\text{BG}$ is a constant background originating from leaked pump light and detector dark counts. 

\begin{figure}[t]
    \centering
    \includegraphics[width=\linewidth]{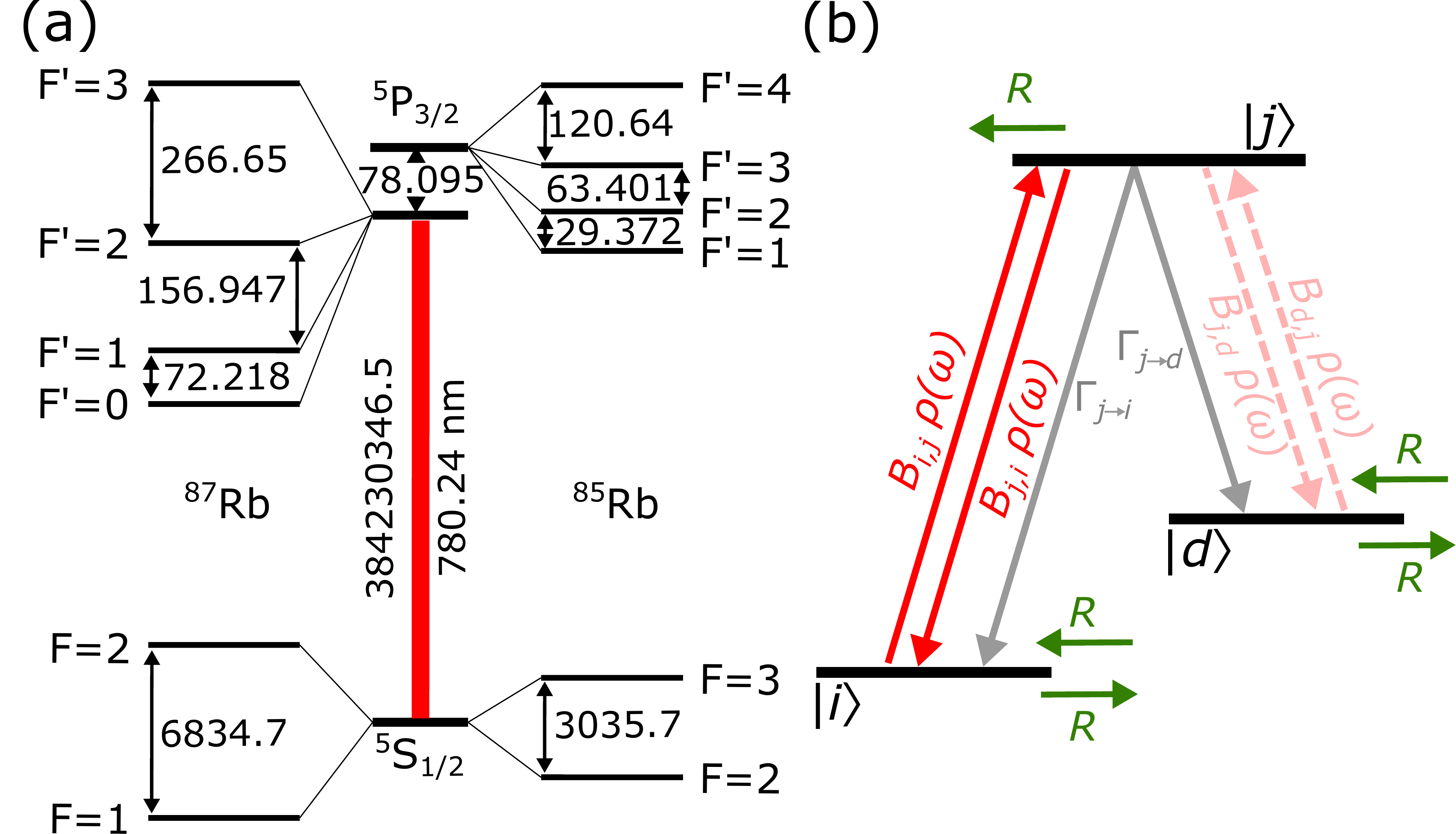}
    \caption{(a) Hyperfine levels of both Rb isotopes for the D$_2$ transition. All values are in MHz unless specified. (b) Three level system used to model the atomic system.  $\ket{i}$ and $\ket{d}$ represent the ground states and $\ket{j}$ the excited state, with the $\ket{i}\rightarrow\ket{j}$ transition being on/near-resonance with the probe and pump lasers. $\ket{d}$ is effectively a dark state, indicated by the dashed arrows.  The steady state populations of each state depend on the transition rates induced by the near-resonant laser fields, $B_{i,j}\rho(\omega)$ and $B_{j,i}\rho(\omega)$, the off-resonant laser fields, $B_{d,j}\rho(\omega)$ and $B_{j,d}\rho(\omega)$, the spontaneous decay rates, $\Gamma_{j \rightarrow i}$ and $\Gamma_{j \rightarrow d}$, and the rate of atomic drift into and out of the beam, $R$.}
    \label{D1D2_3levels}
\end{figure}

The atoms can be treated as a number of open two-level systems \cite{Himsworth2010}. These comprise of a ground state $\ket{i}$ coupled to an excited state $\ket{j}$, which can decay to levels other than $\ket{i}$, represented by a third dark state, $\ket{d}$, shown in Fig. \ref{D1D2_3levels}(b).  The dark state captures decay to all states other than $\ket{i}$. We consider absorption at frequencies close to the $\ket{i} \rightarrow \ket{j}$ transition and will write the absorption cross-section for this reduced system as $\sigma_{i,j}(\omega, \omega_{p})$, which will account for important effects such as optical pumping, saturation and atomic drift.

We will first look to derive $\sigma_{i,j}(\omega, \omega_{p})$ in terms of the Einstein $B$ coefficients %$B_{ij}$ is the coefficient for levels $i$ and $j$, and is given by,
\begin{equation}
    B_{i,j} = \frac{\pi}{\epsilon_{0} \hbar^{2} \mathcal{D}_{i}} \sum_{m_{F}}|\mu_{i,j}|^{2} \, ,
    \label{eq:B}
\end{equation}
\noindent where $\mathcal{D}_{i}$ is the degeneracy of ground level $i$ and $|\mu_{i,j}|=W_{i,j}|\bra{J_{j}|\vec{r}\ket{J_{i}}}=W_{i,j}\mu_{0}$ where $W_{i,j}$ are coefficients calculated from Wigner 3-j and 6-j symbols and $\mu_{0}$ is the reduced transition dipole moment. Details of calculated $B_{i,j}$ values can be found in the Supplemental Material~\cite{supplement}. 

We will also make use of the fractional population $n_{i}$ for energy level $\ket{i}$ and neglect excitation from the dark state $\ket{d}$ due to the large hyperfine ground state splitting.  Off-resonant absorption can be taken into account for the full system by summing over the different ground states. The absorption cross-section can then be written as
\begin{multline}
    \sigma_{i,j}(\omega, \omega_{P}) = \frac{\hbar \omega}{c} \Big(B_{i,j}n_{i}(\omega, \omega_{P})\\ - B_{j,i}n_{j}(\omega, \omega_{P})\Big) L_{i,j}(\omega, v)f(v) \, ,
    \label{eq:sigma1}
\end{multline}
where $\hbar$ is the reduced Planck's constant and $c$ is the speed of light in vacuum. We have also defined a Lorentzian function which characterizes the response of an atomic transition to the incident probe field as
\begin{equation}
    L_{i,j}(\omega, v) = \frac{\Gamma/(2\pi)}{(\omega-\omega_{i,j}-k v)^{2}+(\Gamma/2)^{2}} \, ,
    \label{eq:L}
\end{equation}
where $k=\omega/c$ is the wave vector, $v$ is the velocity of the atoms, and $\Gamma$ is the total excited state decay rate. The term $k v$ accounts for the Doppler shift seen by atoms travelling at velocity $v$.  The function $f(v)$ in Eq.(\ref{eq:sigma1}) is the 1D Maxwell-Boltzmann distribution for the temperature of the vapour cell.  

To find the populations of the different states we solve the rate equations for the open two-level system to find $n_{i}$ and $n_{j}$. A full derivation is presented in the Supplemental Material~\cite{supplement}. This allows us to make the substitution $B_{i,j}n_{i} - B_{j,i}n_{j} = B_{i,j}\mathcal{N}_{i}\Delta N_{i}$ where $\mathcal{N}_{i}$ denotes the initial fractional population of $\ket{i}$ and $\Delta N_{i}$ (related to the population change in $\ket{i}$ and $\ket{j}$ caused by the beams) is defined as
\begin{equation}
    \Delta N_{i}(\omega, \omega_{p}, v) = \frac{1-\sum_{j}\alpha_{i,j}(\omega, \omega_{p}, v)}{1+\sum_{j}\frac{\mathcal{D}_{j}}{\mathcal{D}_{i}}\alpha_{i,j}(\omega, \omega_{p}, v)\beta_{i,j}} \, ,
    \label{eq:deltaNi}
\end{equation}
where we have included multiple excited levels by including a sum over all excited states $\ket{j}$. Here we have defined a saturation parameter
\begin{align}
    \alpha_{i,j}(\omega, \omega_{p}, v) &= \frac{B_{j,i}\rho_{i,j}(\omega, \omega_{p}, v)}{ B_{j,i}\rho_{i,j}(\omega, \omega_{p}, v)+\Gamma + R_{i,j}} \, ,
    \label{eq:alphaij}
\end{align}
and an optical pumping parameter
\begin{equation}
    \beta_{i,j} = 1 + \frac{\Gamma - \Gamma_{j\rightarrow i}}{R_{i,j}} \, ,
    \label{eq:beta}
\end{equation}
where $\Gamma_{j\rightarrow i}$ is the decay rate from excited state $\ket{j}$ to ground state $\ket{i}$ and $R_{ij}$ is the rate at which atoms enter and leave the beam (see Fig. \ref{D1D2_3levels} (b)), both of which depend on the transition being considered~\cite{supplement}. The spectral energy density for the $\ket{i}\rightarrow\ket{j}$ transition is
\begin{equation}
    \rho_{i,j}(\omega, \omega_{p}, v) = L_{i,j}(\omega, v)I_{0}/c + L_{i,j}(\omega_{p}, v)I_{P}/c
    \label{eq:rho}
\end{equation}
where $I_{P}$ as the pump laser intensity. The resulting absorption cross-section for the full system is
\begin{multline}
    \sigma(\omega,\omega_p) = \frac{-\hbar \omega}{c} \int_{-\infty}^{\infty} \sum_{i,j} B_{i,j}\mathcal{N}_{i}\Delta N_{i}(\omega, \omega_{p}, v)\\
    \times L_{i,j}(\omega, v) f(v)  dv \, ,
    \label{eq:sigma3}
\end{multline}
where the summation is over all ground and excited states.

Finally the fluorescence term in Eq.(\ref{eq:I}) is given by,
\begin{multline}
    I_\text{fluo}(\omega_{p},v) = \frac{N_{v}}{3}\eta_\text{fluo}\Gamma\pi r^{2} l \\ \times \Big(\int_{-\infty}^{\infty} f(v)
    \sum_{i,j}\frac{\alpha_{i,j}(\omega, \omega_{p}, v)}{1+\frac{g_{j}}{g_{i}}\alpha_{i,j}(\omega, \omega_{p}, v)\beta_{i,j}}dv\Big) \, .
    \label{eq:F}
\end{multline}

\begin{figure*}[t]
    \includegraphics[width=0.96\linewidth,clip]{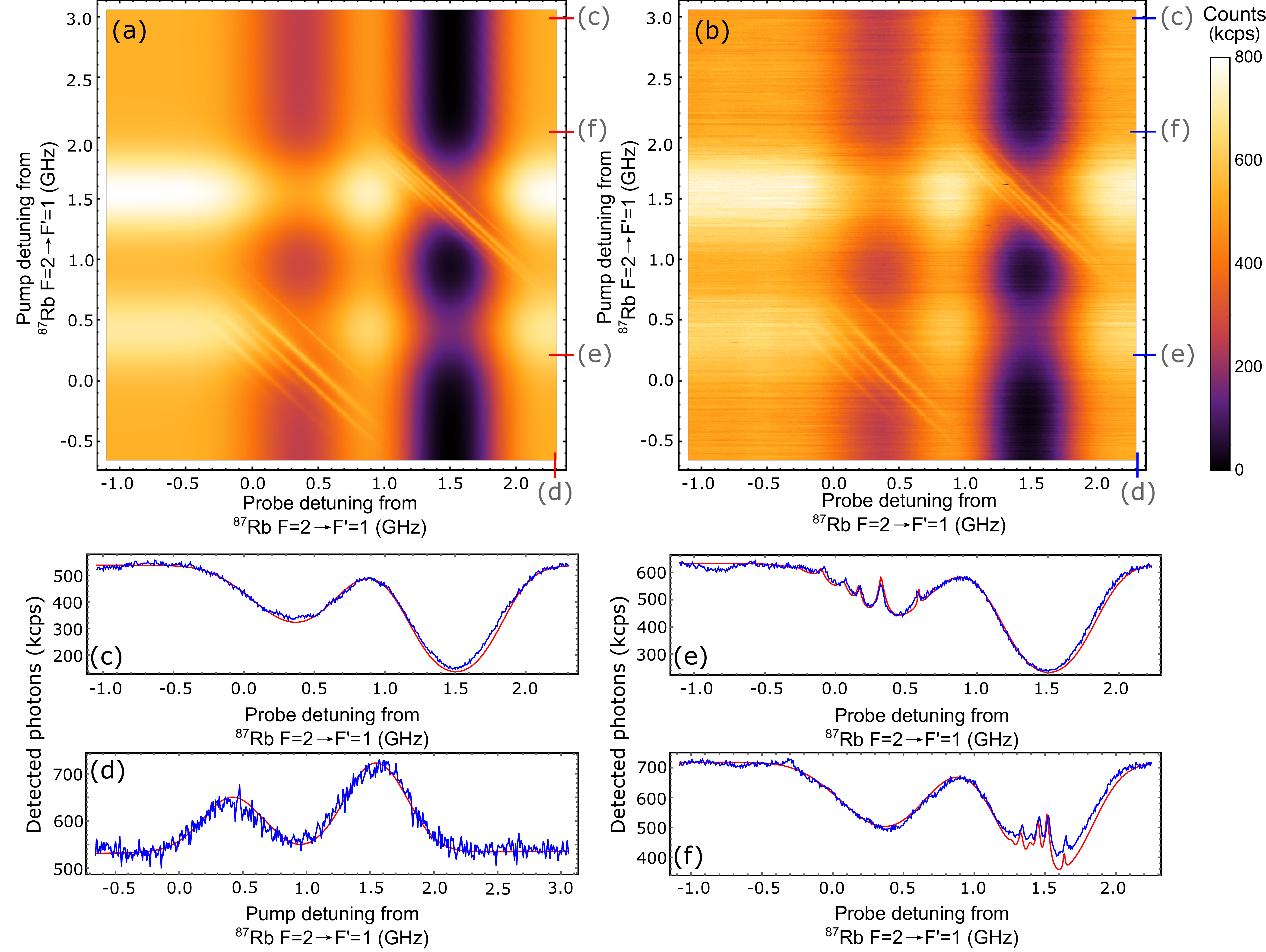}
    \caption{Density plots of the number of transmitted probe photons through the Rb cell with varying probe laser frequency on the x-axis and varying pump laser frequency on the y-axis. The anti-diagonal sub-Doppler features towards the bottom left result from the $^{87}$Rb D2 F$=2\rightarrow $F'$=1,2,3$ transitions and the features towards the top right result from $^{85}$Rb D2 F$=3\rightarrow $F'$=2,3,4$ transitions. (a) Simulation. (b) Experimental data.  Cut-throughs of the simulated (red) and experimental (blue) density plots, with the positions of the cut-throughs indicated on the density plots. (c) Horizontal slice showing the room temperature Doppler broadened spectra. (d) Vertical slice showing the fluorescence produced by the pump beam. (e) Horizontal slice showing the sub-Doppler features of $^{87}$Rb. (f) Horizontal slice showing the sub-Doppler features of $^{85}$Rb.} 
    \label{data}
\end{figure*}

The factor inside the summation is the steady state population in the $^{5}P_{3/2}$ excited states and $r$ is the radius of the pump beam.  In addition to the summations shown in Eq.(\ref{eq:sigma3}) and Eq.(\ref{eq:F}), there is a further summation over the two Rb isotopes.  We performed the integral in Eq.(\ref{eq:sigma3}) and Eq.(\ref{eq:F}) numerically for 800 velocity classes, where the velocities considered were those within three standard deviations of the mean (equal to 0 in this 1D case), with more velocity classes being sampled at a closer proximity to the mean.  At each velocity, all energy levels were considered in order to account for off-resonant pumping. We accounted for a diverging pump beam radius by spatially dividing the vapour cell into 10 slices and propagating the output from each slice into the next. We found the total collection and detection efficiency of the probe to be $\eta_\text{probe} = 0.215$ and of the fluoresence to be $\eta_\text{fluo}=1.13 \times 10^{-9}$, demonstrating that the polarization and spatial filtering, as well as the extra propagation distance between the cell and the fiber, allowed for a high rejection of fluorescence.

\subsection{Results and Discussion}
The result of the calculation of Eq.\ref{eq:I} is shown in Fig.\,\ref{data}(a), while the measured data is shown in Fig.\,\ref{data}(b). The probe frequencies in the experimental data were corrected for slow laser drifts using the Doppler-broadened absorption features in each scan, and the pump frequencies were corrected to ensure parallel sub-Doppler features separated according to the known hyperfine transition frequencies \cite{Siddons2008a}. The simulated and measured spectra show a high level of agreement, having a mean absolute percentage error (MAPE) of 4.17\%~\cite{supplement}.

When the probe frequency is varied and the pump is far off-resonance with any rubidium transitions, we observe Doppler-broadened absorption dips, shown in Fig.\,\ref{data}(c).  Fluorescence peaks are seen in Fig.\,\ref{data}(d) for constant probe frequency as the pump laser is tuned across the Doppler broadened features.

When the probe and pump frequencies approach resonance, the broad absorption and fluorescence features intersect. Within these areas there are probe and pump frequencies that are similar enough that the Doppler shift due to the motion of the atoms compensates for the difference in frequency. This means that sub-Doppler features appear along anti-diagonal lines in the two-dimensional spectrum. Along these anti-diagonal lines the two lasers are addressing different narrow velocity classes of the atomic motion. When the lasers are exactly resonant they are addressing the atoms with zero velocity and this would be a diagonal line through the spectrum. Tracing this diagonal recovers the usual saturated absorption spectrum, where only one laser is used for both pump and probe \cite{Himsworth2010, Preston1996}. In Fig.\,\ref{data}(e) and (f) we can see the cases for constant pump frequencies around the $^{87}$Rb and $^{85}$Rb transitions, which show sub-Doppler features and allow us to visualise any differences between the simulated and measured data. 

There are a number of faint narrow diagonal features in the measured two-dimensional spectrum. These were caused by reflections of the pump beam from the faces of the optical elements of the setup, leading to a faint pump beam co-propagating with the probe beam. The two beams see atoms Doppler-shifted in the same direction. These features are off of the diagonal and cannot be observed in a normal sub-Doppler spectrum with only one laser.  These off-diagonal features are not included in the simulated spectrum and so are a small contribution to the quoted error between the simulation and data.

We largely see close agreement, with the discrepancies most likely resulting from imperfect pump and probe laser frequency correction, a variation in fiber coupling before detection caused by mechanical vibrations of the set up, and any error in our measurement of the average cell temperature. The large fluctuations in the vertical cut-through (Fig.\,\ref{data}(d)) are a result of the longer time-scales being considered in this direction, and are a main contributor to the quoted MAPE.  We can also see that the simulation underestimates the intensity of some of the sub-Doppler peaks.  We believe this is due to our model failing to accurately reproduce the optical pumping behaviour arising from the diverging pump beam.
%To achieve the demonstrated agreement we had to adjust several parameters that we used as the input to the simulation. The fluorescence was minimised by collecting the transmitted light a large distance from the cell but we still observe this fluorescence in the measured spectrum. We took the cut-through in fig. \ref{data}e to fit the amplitude of the fluorescence. The beam radius was found by taking cuts-through at ?? \ref{data}d & \ref{data}f. Once these values were found for single cuts-through we then simulated the entire spectrum.
%
%The effect of varying the pump intensity in each slice as seen in fig. \ref{slices} was not very significant but the fact that this also allowed the beam radius to change accounted allowed us to get better agreement between the simulation and the measured spectrum.  
%How much of an effect does this actually have ??

\section{\label{sec:level4}Conclusion}
We have measured a pump-probe spectrum of rubidium, showing the interaction of a single-photon-level probe beam with hyperfine levels in the presence of a pump laser that can be independently tuned in frequency. We have compared this spectrum to a rate-equation model and have found good agreement. This model is valid for any gas at standard pressures where coherent and non-linear effects can be discounted. This spectrum allows us to observe the sub-Doppler features for non-zero velocity classes in the atomic vapour, and also shows faint effects of a co-propagating pump, neither of which can be observed with standard saturated absorption spectroscopy. 
As in the standard saturated absorption spectra, optical pumping has a significant effect.  The main difference in the single-photon-level regime is the increase in background due to fluorescence. A good understanding of the absorption spectra is important for determining the temperature, number density, population of occupied levels or the chemical composition of the medium \cite{Corney1977, Eckbreth2019}, as well as standard applications of atomic vapours such as laser locking \cite{Mccarron2016, Genov2017}, compact magnetometry \cite{Schwindt2004}, and atomic quantum memories \cite{Heshami2016}. 
This setup is ideal for replacing the low intensity probe laser with an appropriate single photon source. Single molecules of dibenzoterrylene, an aromatic hydrocarbon, are an ideal candidate with emission that is resonant with the D2 absorption lines of Rb \cite{Major2015, Pazzagli2018} that has been shown to be Stark tuned by 100's of GHz \cite{Schadler2019}. The interfacing of single photons and the internal states of atomic vapours is highly desirable for building new quantum technologies.

\section*{\label{sec:level5}Acknowledgements}
We are grateful to Ed Hinds for insightful discussions and thank Jon Dyne and Dave Pitman for their expert mechanical workshop support. This work was supported by EPSRC (EP/P030130/1, EP/P01058X/1, and EP/R044031/1), the Royal Society (UF160475, RGF/R1/180066, and RGF/EA/180203), and the EraNET Cofund Initiative QuantERA under the European Union’s Horizon 2020 research and innovation programme, Grant No. 731473 (ORQUID Project).

\bibliography{bibliography}% Produces the bibliography via BibTeX.

\newpage
\clearpage

\section*{\label{Appendix}Supplemental Material}
\subsection{Rate equations and solutions for a three-level system}

Here we present a more complete derivation of the three-level rate equation solution, used to derive Eq.(\ref{eq:sigma3}).  We explicitly omit the dependent variables until the end for legibility. For our three-level system, the fractional population of each level can be written in the form of rate equations.  Here we define the fractional population of the ground state as $n_{i}$, the coupled excited state as $n_{j}$ and a third `dark' state as $n_{d}$. The fractional populations for the three levels are given by,
\begin{align}
    &\begin{aligned}
    \dot{n}_{i} =
         -n_{i}[B_{i,j}\rho_{i,j} + R_{i,j}] + n_{j}&[B_{j,i}\rho_{i,j} + \Gamma_{j\rightarrow i}]\\ &+ \mathcal{N}_{i}R_{i,j}
    \end{aligned}\\
    &\begin{aligned}
    \dot{n}_{j} =
         n_{i}B_{i,j}\rho_{i,j} - n_{j}[B_{j,i}\rho_{i,j} + &B_{j,d}\rho_{d,j} \Gamma_{j\rightarrow i} + R_{i,j}] \\&+ n_{d}B_{d,j}\rho_{j,d}
    \end{aligned}\\
    &\begin{aligned}
    \dot{n}_{d} =
         -n_{d}[B_{d,j}\rho_{d,j} + R_{d,j}] + n_{j}&[B_{j,d}\rho_{d,j} + \Gamma_{j\rightarrow d}]\\ &+ \mathcal{N}_{d}R_{d,j}
    \end{aligned}
\end{align}
where $\rho_{i,j}$ is the spectral energy density for the $\ket{i}\rightarrow\ket{j}$ transition, given by,
\begin{equation}
    \rho_{i,j} =
    L_{i,j}\frac{I_{0}}{c} + L_{i,j}\frac{I_{P}}{c}
\end{equation}
with $\omega_{0}\,(\omega_{P})$ and $I_{0}\,(I_{P})$, the probe (pump) laser angular frequency and intensity respectively. The initial fractional population in ground states $\ket{i}$ and $\ket{d}$ are
\begin{align}
    \mathcal{N}_{i} &= \frac{\mathcal{D}_{i}}{\mathcal{D}_{i} + \mathcal{D}_{d}}\\
    \mathcal{N}_{d} &= \frac{\mathcal{D}_{d}}{\mathcal{D}_{i} + \mathcal{D}_{d}}
\end{align}
with $\mathcal{D}_{i}$ representing the degeneracy in state $\ket{i}$ and $\mathcal{D}_{d}$ representing the degeneracy in state $\ket{d}$. The initial populations are assumed to be housed solely in the two ground states of the system (i.e. $\mathcal{N}_{i} + \mathcal{N}_{d} = 1$) since the thermal energies are on the order of THz and the excited state is several hundred THz away, for our system.

Now assuming the system has reached a steady state ($\dot{n}_{i}=\dot{n}_{j}=\dot{n}_{d}=0$), we arrive at,
\begin{align}
    n_{i} &= \frac{n_{j}[B_{j,i}\rho_{i,j} + \Gamma_{j \rightarrow i}] + \mathcal{N}_{j}R}{B_{j,i}\rho_{i,j} + R_{i,j}}
    \label{ng}\\
    n_{j} &= \frac{n_{i}B_{i,j}\rho_{i,j} + n_{d}B_{d,j}\rho_{d,j}}{B_{j,i}\rho_{i,j} + B_{j,d}\rho_{j,d} + \Gamma + R_{i,j}}
    \label{ne}\\
    n_{d} &= \frac{n_{j}[B_{j,d}\rho_{j,d} + \Gamma_{j \rightarrow d}] + \mathcal{N}_{d}R_{d,j}}{B_{d,j}\rho_{d,j} + R_{d,j}}
    \label{nd}
\end{align}
Substituting Eq.(\ref{ne}) into Eq.(\ref{ng}) results in,
\begin{equation}
    n_{i} =  \frac{\mathcal{N}_{i}}{1+\frac{\mathcal{D}_{j}}{\mathcal{D}_{i}}\alpha_{i,j}\beta_{i,j}}
    \label{fullng}
\end{equation}
with the saturation parameter $\alpha_{ij}$ defined as,
\begin{align}
    \alpha_{i,j} &= \frac{B_{j,i}\rho_{i,j}}{ B_{j,i}\rho_{i,j}+\Gamma + R_{i,j}}%\\
    %&= \Big(1+\frac{\Gamma_{2} + R}{B_{21}L(\omega_{0}, \omega_{P},v)I/c}\Big)^{-1}
\end{align}
and the optical pumping parameter $\beta_{i,j}$ defined as,
\begin{equation}
    \beta_{i,j} = 1 + \frac{\Gamma - \Gamma_{j\rightarrow i}}{R_{i,j}}
\end{equation}
In deriving Eq.(\ref{fullng}), we have assumed that no considerable population from $n_{d}$ is transferred to $n_{j}$, due to $\rho_{d,j}\ll\rho_{i,j}$.  However, when extending to multiple levels and summing over all levels, absorption due to $\rho_{d,j}$ is explicitly taken into account, allowing this model to account for power broadening.  This derivation is valid for any open two-level system, and hence can be readily extended to a multilevel system by summing over many three-level systems.

Perhaps a more intuitive understanding of $\alpha_{i,j}$ can be gained by combining it with the degeneracy term preceding it in Eq.(\ref{fullng}), using the identity $\frac{\mathcal{D}_{j}}{\mathcal{D}_{i}}B_{j,i} = B_{i,j}$,
\begin{equation}
    \frac{\mathcal{D}_{j}}{\mathcal{D}_{i}}\alpha_{i,j} = \frac{B_{i,j}\rho_{i,j}}{B_{j,i}\rho_{i,j} + \Gamma + R_{i,j}}
\end{equation}
leading to a term pertaining to the ratio of the transition rate from state $\ket{i}$ to $\ket{j}$ and all transitions removing population from state $\ket{j}$.  

The optical pumping parameter, $\beta_{i,j}$, characterizes population lost to dark states, normalized by the rate of atoms drifting into the beam.

We return to the term $(B_{i,j}n_{i} - B_{j,i}n_{j})$ in Eq.(\ref{eq:sigma1}), which we can now derive an expression for in terms of $\mathcal{N}_{i}$,
\begin{equation}
    B_{i,j}n_{i} - B_{j,i}n_{j} = B_{i,j}\mathcal{N}_{i}\Delta N_{i}
\end{equation}
where $\Delta N_{i}$ is defined as,
\begin{equation}
    \Delta N_{i} = \frac{1-\alpha_{i,j}}{1+\frac{\mathcal{D}_{j}}{\mathcal{D}_{i}}\alpha_{i,j}\beta_{i,j}}
    \label{deltaNg}
\end{equation}
and is related to the difference in population of levels $\ket{i}$ and $\ket{j}$ caused by the probe and pump beams.  Here we have used Eq.(\ref{fullng}) and a similar equation derived for $n_{j}$.  The numerator in Eq.(\ref{deltaNg}) accounts for population pumped into the excited state from the ground states, while the denominator accounts for the populations transitioning to the two ground states.  Due to the presence of state $\ket{d}$ and atomic motion into and out of the beam, not all of the population excited to state $\ket{j}$ returns to state $\ket{i}$ - parameterized by $\beta_{i,j}$.

Putting it all together, the final $\sigma(\omega_{0}, \omega_{P})$ for an open two-level system is,
\begin{equation}
    \begin{aligned}
    \sigma(\omega_{0}, \omega_{P}) = \frac{-\hbar \omega_{0}}{c} \int_{-\infty}^{\infty} B_{i,j}\mathcal{N}_{i}\Delta N_{i}(\omega_{0}, \omega_{P}, v)\\ \times L_{i,j}(\omega_{0}, \omega_{P}, v) f(v)  dv \, .
    \end{aligned}
    \label{sigma2level}
\end{equation}
Through adding the relevant sums to Eqs.(\ref{deltaNg}, \ref{sigma2level}), as in the main paper, we can model a system with an arbitrary number of levels. 

\subsection{\label{subsection:R}Deriving $R_{i,j}$}

Due to the large effect optical pumping has on the system, we require an accurate determination of the rate at which atoms in particular states enter and leave the beam, $R_{i,j}$.  The smaller $R_{i,j}$, the larger the amount of optical pumping in the system. 

We can initially approximate the rate at which atoms traverse the beam as,
\begin{equation}
    R_{0} = \frac{v_{p}}{2r} \, ,
\end{equation}
where $v_{p}=\sqrt{2k_{B}T/M}$ is the most probable speed of an atom in 1D and $r$ is the $1/e^{2}$ radius of the pump beam \cite{Demtroder}.

However, an atom may be pumped to a dark state with the absorption of a single photon and so optical pumping can continue well outside the the $1/e^{2}$ radius of the beam.  As in Himsworth et al. \cite{Himsworth2010}, we account for this by scaling the radius, $r$, of the pump beam dependent on the transition begin addressed.  If a transition has a higher transition strength, a larger beam radius is considered, as an atom is more likely to be pumped to a dark state in the wings of the laser.

We quantify the intensity at which optical pumping stops through a parameter proportional to the reduced saturation intensity, given by\cite{Demtroder},
\begin{equation}
    I^{R}_{i,j} = \frac{I_{i,j}}{\Gamma}\Big(\frac{\Gamma+R_{0}}{2 + \frac{(\Gamma-\Gamma_{j\rightarrow i })}{R_{0}}}\Big)\, ,
\end{equation}
where $I_{i,j}$ denotes the saturation intensity when treating the $i\rightarrow j$ transition as a two level system.  For details on calculating $I_{i,j}$ see Section \ref{params}.  

We then define $R$ such that,
\begin{equation}
    R_{i,j} = \frac{R_{0}}{\sqrt{\frac{1}{2}ln(\frac{I_{P}}{I^{R}_{i,j}})}}\, , 
\end{equation}
which has the effect of scaling the radius considered to the point where the intensity of the pump beam equals $I^{R}_{i,j}$.

The final part to take into account is the varying radius of the diverging pump beam through the cell.  This results in a position-dependent optical pumping.  To include this detail, we first split the Rb vapour cell into a number of slices and determine the pump power and radius entering each of the consecutive slices.  This allows us to compute $R_{i,j}$ for each slice.  The probe is then allowed to propagate through each slice in turn (from the opposite direction), with a different $R_{i,j}$ and pump intensity present in each slice.  The output of each slice is fed forward into the next, thus allowing the previous theoretical derivation to be applied to a spatially varying pump laser.

\subsection{1D Maxwell Boltzmann Distribution}

We have taken the velocity distribution $f(v)$, to be a 1D Maxwell Boltzmann distribution, as we are only considering velocities, $v$, along the longitudinal axes of the laser fields,
\begin{equation}
    f(v) = \sqrt{\frac{M}{2\pi k_{B}T}}e^{- \frac{M v^{2}}{2k_{B}T}}
\end{equation}
with $M$ the mass of an atom, $k_{B}$ Boltzmann's constant, and $T$ the temperature of the vapour cell.

\subsection{\label{params}Simulation parameters}

\begin{table*}
\renewcommand{\arraystretch}{1.4}
\begin{tabular*}{\textwidth}{c|| c c c || c c}
\hline \hline
   & ~~~~~$i\rightarrow j$~~~~~ & ~~~~~$\Gamma_{j\rightarrow i}$~~~~~ & ~~~~~$\mu_{i,j}$~~~~~ & ~~~$B_{i,j}$
             (s$^{-1}\cdot $[Jm$^{-3}$Hz$^{-1}$]$^{-1}$ $\times 10^{20}$)~~~~~ &
             ~~~~~$I_{i,j}$ (Wm$^{-2}$)~~~~~
              \\
 \hline \hline
 ~~~~$^{87}$Rb~~~~ & \begin{tabular}[c]{c@{}}$2\rightarrow 1$ \\
            $2 \rightarrow 2$ \\
            $2 \rightarrow 3$ \\
            $3 \rightarrow 2$ \\
            $3 \rightarrow 3$ \\
            $3 \rightarrow 4$ \\
            \end{tabular}            
      & \begin{tabular}[c]{c@{}}$1$ \\
            $5/6$ \\
            $1/2$\\
            $1/6$ \\
            $1/2$ \\
            $1$ \\
            \end{tabular} 
      & \begin{tabular}[c]{c@{}}$1/6$ \\
            $5/12$ \\
            $5/12$ \\
            $1/12$ \\
            $5/12$ \\
            $7/6$ \\
            \end{tabular}
      & \begin{tabular}[c]{c@{}}$7.5880$ \\
            $18.9699$ \\
            $18.9699$ \\
            $2.2765$ \\
            $11.3825$ \\
            $31.8710$ \\
            \end{tabular}       
      & \begin{tabular}[c]{c@{}}$200.2738$\\
            $80.1096$ \\
            $80.1097$ \\
            $400.5264$ \\
            $80.1054$ \\
            $28.6091$ \\
            \end{tabular} \\
\hline
$^{85}$Rb & \begin{tabular}[c]{c@{}}$1 \rightarrow 0$ \\
            $1 \rightarrow 1$ \\
            $1 \rightarrow 2$ \\
            $2 \rightarrow 1$ \\
            $2 \rightarrow 2$ \\
            $2 \rightarrow 3$ \\
            \end{tabular}  
     & \begin{tabular}[c]{c@{}}$1$ \\
            $7/9$ \\
            $4/9$ \\
            $2/9$ \\
            $5/9$ \\
            $1$ \\
            \end{tabular}       
      & \begin{tabular}[c]{c@{}}$1/2$ \\
            $35/54$ \\
            $14/27$ \\
            $5/27$ \\
            $35/54$ \\
            $3/2$ \\
            \end{tabular}
      & \begin{tabular}[c]{c@{}}$13.6622$ \\
            $17.7102$ \\
            $14.1681$ \\
            $3.6144$ \\
            $12.6505$ \\
            $29.2767$ \\
            \end{tabular}       
      & \begin{tabular}[c]{c@{}}$66.7743$ \\
            $51.5116$ \\
            $64.3895$ \\
            $180.2863$ \\
            $51.5105$ \\
            $22.2576$ \\
                \end{tabular}  \\
\hline
\end{tabular*}
\caption{Key parameters for $^{87}$Rb and $^{85}$Rb D$_2$ transition.  Here $i$ represents the ground state and $j$ the excited state.  The decay rate of the transition $\ket{i}\rightarrow\ket{j}$, $\Gamma_{j\rightarrow i}$ is given as a fraction of the total excited state decay rate, $2\pi \times 6.065$\,MHz and $2\pi \times 6.0666$\,MHz for $^{87}$Rb and $^{85}$Rb respectively. The transition dipole moments, $\mu_{ij}$ are given as ratios of the reduced dipole moment $\mu_{0} = \bra{J_{j}}|\vec{r}|\ket{J_{i}}$.  Both of these parameters are input into the simulation.  The B-coefficients, $B_{i,j}$ and saturation intensities $I_{i,j}$ are computed from the simulation, using Eq.(\ref{eq:B}) and Eq.(\ref{Isat}) respectively.}
\label{ParamTable}
\end{table*}

Here we present the input parameters used by the simulation, as well as those computed by the simulation. The Rb cell used was 7.5\,cm long and contained the natural abundances of $0.2785$ and $0.7215$ for $^{87}$Rb and $^{85}$Rb respectively.  The temperature was held constant at $40^{\circ}$C and the number densities for $^{87}$Rb and $^{85}$Rb were determined by the simulation to be $1.1739404423746728\times 10^{16}$\,m$^{-3}$ and $3.0443148302615932\times 10^{16}$\,m$^{-3}$ respectively, from the relation,
\begin{equation}
    ^{\mathcal{I}}N_{v} = \frac{\mathcal{P}_{\mathcal{I}}}{k_{B}T} \times 133.323 \times 10^{p_{y}} \,
\end{equation}
where $\mathcal{I}$ denotes the isotope, $\mathcal{P}$ is the isotope fraction, $k_{B}$ is the Boltzmann constant and $T$ is the temperature. The term $p_{y}$ represents the vapour pressure and is determined by,
\begin{equation}
    p_{y} = \begin{cases}
        -94.04825 - (\frac{1961.258}{T}) & T <39.3^{o}\text{C} \\
        - 0.03771678 T + 42.57526 \log_{10}(T), & \\\\
    15.88253 - (\frac{4529.535}{T}) & T \geq 39.3^{o}\text{C,} \\
    + 0.00058663 T - 2.99138 \log_{10}(T), &
    \end{cases}
\end{equation}
relating to the melting point of $39.3^{o}$C for Rb.

The excited state total decay rates for $^{87}$Rb and $^{85}$Rb are $2\pi \times 6.065$\,MHz and $2\pi \times 6.0666$\,MHz.  The decay rates for the individual transitions are given as a fraction of this total decay rate, in Table \ref{ParamTable}.  Also given in Table \ref{ParamTable} are the transition dipole strengths, $\mu_{ij}$ for the transition $i\rightarrow j$ in units of the reduced dipole moment,  $\mu_{0} = \bra{J_{j}}|\vec{r}|\ket{J_{i}}$.  These are equivalent to the $W_{ij}$ coefficients calculated from Wigner 3-j and 6-j symbols, referred to in the main text.

The $B$-coefficients were computed according to Eq.(\ref{eq:B}) and are shown in Table \ref{ParamTable}, along with the saturation intensities for each of the transitions.  Here we define saturation intensity in the usual way, as the intensity needed for the population difference of a two-level system to equal one half.  It can be shown that this is equivalent to \cite{Foot2007},
\begin{equation}
    I_{i,j} = \frac{1}{3} \frac{\pi c \Gamma^{2}}{4 \mathcal{D}_{j} B_{j,i}} ,
    \label{Isat}
\end{equation}
where $c$ is the speed of light in vacuum,  $\Gamma$ is the total decay rate and $D_{j}$ is the degeneracy of excited state $\ket{j}$.  We have pulled out the factor of a third to make it clear we are considering linearly polarised light.

As for the laser properties, the power of the pump beam was $790\,\mu$W.  The number of photons arriving at the cell was determined by first measuring the fraction of power coupled through the set up with a relatively high power laser, off resonance.  Using this fraction, we could then determine the number of photons arriving at the cell, given the number of photons collected.  We determine the number of photons arriving at the cell to be $2.5(1)\times 10^{6}$ photons/s and set $\eta_{probe} = 0.215$.  The background was measured to be $\sim10000$\,counts/s, which includes contributions from room lights, unfiltered pump laser and dark counts. The radius of the pump beam was set to be diverging through the cell and was determined through fitting the simulation to the data, as was $\eta_{fluo}$.

\subsection{Accounting for laser frequency drift}

Throughout the full two-dimensional scan, while both pump and probe lasers were locked to external cavities, some drift from the intended frequency scans was observed.

The minima of the Doppler broadened spectra varied throughout the scan, which indicated inconsistency in the probe frequency scanning.  To correct for this, we fitted both Doppler features using Gaussian functions with negative amplitudes.  To avoid the sub-Doppler features causing an offset in the fitted minima, we used only the Doppler dips absent of sub-Doppler features for fitting.  From the computed minima, we could remove any offset and re-scale the probe frequencies.

After correcting for probe drift between scans, we corrected for pump laser drift.  We began by assuming that the pump laser remained at a constant frequency during each individual probe scan. Plotting the anti-diagonal sub-Doppler features, we observed a deviation from the expected linear dependence.  By re-scaling the pump axis using a quadratic function to straighten the sub-Doppler features, as well as scale the separation between fluorescent peaks and Doppler absorption dips, we correct the full two-dimensional scan to be square, meaning both the pump and probe axes are within a single factor of the true frequency scales, though with different absolute offsets.

We can deduce this scale factor by extracting a diagonal from the two-dimensional plot and finding the frequency separations of the hyperfine peaks.  The diagonal is related to the usual saturated absorption spectrum, though the intensity of the features is modified by the fluorescence and there may be a relative constant offset between the pump and probe frequencies, resulting in a change in the absolute positions of the peaks, but not their separation.  The diagonal is found by comparing the minimum of a sub-Doppler dip and the peak of the corresponding fluorescence, giving the required $(x, y)$ coordinate.  Here we used the $^{87}$Rb isotope peak and dip.

After picking out the hyperfine peaks from the data, we plot the frequency locations of these peaks against the values quoted in the literature\cite{Steck2003, Steck2013} and fit a quadratic function to minimize the difference between the two lists of frequencies.  After applying this scaling, the probe and pump axes are within an offset of the accurate frequencies.  This final offset is then obtained by comparing the simulation to the measured data.

\subsection{\label{Sec:nd}Quantifying the difference between simulation and data}

We can quantify the difference between two matrices (in this case the pixels of two images) by calculating the mean absolute percentage error (MAPE) defined as

%\begin{equation}
%    nd = \frac{\sum_{ik} |M_{ik} - N_{ik}|}{\sum_{ik} \sqrt{M_{ik}N_{ik}}}
%\end{equation}

\begin{equation}
    \text{MAPE} = \frac{1}{N} \sum_{a,b} \left| \frac{S_{a,b} - M_{a,b}}{S_{a,b}} \right|
\end{equation}

\noindent where $S$ and $M$ represent the simulated and measured two-dimensional datasets respectively.  The subscripts $a$ and $b$ represent the elements of each matrix, which are simply the values of the pixels of density plots shown in Fig.~\ref{data}(a) and Fig.~\ref{data}(b).

\end{document}